\documentclass[prb,twocolumn]{revtex4}%
\usepackage{color}

\newcommand{\be}{\begin{equation}}
\newcommand{\ee}{\end{equation}}
\newcommand{\bea}{\begin{eqnarray*}}
\newcommand{\eea}{\end{eqnarray*}}
\newcommand{\bean}{\begin{eqnarray}}
\newcommand{\eean}{\end{eqnarray}}

\begin{document}

\draft
\title
{\bf Effects of interdot hopping and Coulomb blockade on the thermoelectric properties of serially coupled quantum dots}

\author{ David M.-T. Kuo$^{1,\dagger}$ and Yia-Chung Chang$^{2,*}$}
\address{$^{1}$Department of Electrical Engineering, and department of physics, National Central
University, Chungli, 320 Taiwan}

\address{$^{2}$Research Center for Applied Sciences, Academic
Sinica, Taipei, 115 Taiwan}

\begin{abstract}
We have theoretically studied the thermoelectric properties of
serially coupled quantum dots (SCQDs) embedded in an insulator
connected to metallic electrodes. In the framework of Keldysh
Green's function technique, the Landauer formula of transmission
factor is obtained by using the equation of motion method. Based on
such analytical expressions of charge and heat currents, we
calculate the electrical conductance, Seebeck coefficient, electron
thermal conductance and figure of merit ($ZT$) of SCQDs in the
linear response regime. The effects of interdot hopping and electron
Coulomb interactions on $ZT$ are analyzed. We demonstrate that $ZT$
is not a monotonic increasing function of interdot electron hopping
strength ($t_c$). We also show that in the absence of phonon thermal
conductance, SCQD can reach the Carnot efficiency as $t_c$
approaches zero.
\end{abstract}

\date{\today}
\maketitle



\textbf{Introduction}

Recently, many considerable studies have been devoted to seeking
efficient thermoelectric materials with the figure of merit ($ZT$)
larger than 3 because there exist potential applications of solid
state thermal devices such as coolers and power generators.[1-6]
Some theoretical efforts have pointed out that a single quantum dot
(QD) junction system can have a very impressive $ZT$ in the absence
of phonon conductance.[7-9] However, in practice it is difficult to
maintain a large temperature gradient needed to produce sufficient
temperature difference across the nanoscale junction. To reduce the
temperature gradient across the QD junction, it is essential to
consider many serially coupled QDs.[1,5] The transport property of a
junction involving N serially coupled QDs with strong electron
Coulomb interactions is one of the most challenging topics of
condensed matter physics. To gain some insight, we investigate in
the present paper the thermoelectric effect of serially coupled
quantum dots (SCQD) as shown in the inset of Figure 1(a).

It has been shown that the transport properties of the SCQD system
exhibit several interesting features, including current
rectification (due to the Pauli spin blockade), negative
differential conductance, nonthermal broadening of tunneling
current, and coherent tunneling in the Coulomb blockade regime.[10]
Although many theoretical investigations of the above phenomena have
been reported, most of them did not investigate the thermoelectric
properties of SCQDs.[11-13] This study investigates the ZT of SCQD
embedded in a semiconductor nanowire with small phonon thermal
conductance.[4] It is expected that SCQD system has a potential to
enhance the ZT of nanowires. Here we consider nanoscale
semiconductor QDs, in which the energy level separations are much
larger than their on-site Coulomb interactions and thermal energies.
Thus, only one energy level for each quantum dot needs to be
considered. A two-level Anderson model[13] is employed to simulate
the SCQD junction system.


\textbf{Theoretical model}

Using the Keldysh-Green's function technique,[13] the charge and
heat currents of SCQD connected to metallic electrodes are given by
\begin{eqnarray}
J&=&\frac{2e}{h}\int d\epsilon {\cal T}(\epsilon)
[f_L(\epsilon)-f_R(\epsilon)],\\ Q&=& \frac{2}{h}\int d\epsilon
{\cal T}(\epsilon)(\epsilon-E_F-e\Delta V)
[f_L(\epsilon)-f_R(\epsilon)]
\end{eqnarray}
where ${\cal T}(\epsilon)\equiv ({\cal T}_{12}(\epsilon) +{\cal
T}_{21}(\epsilon))/2$ is the transmission factor.
$f_{L=1(R=2)}(\epsilon)=1/[e^{(\epsilon-\mu_{L(R)})/k_BT_{L(R)}}+1]$
denotes the Fermi distribution function for the left (right)
electrode. The left (right) chemical potential is given by
$\mu_L(\mu_R)$. $T_{L(R)}$ denotes the equilibrium temperature of
the left (right) electrode. $e$ and $h$ denote the electron charge
and Planck's constant, respectively. ${\cal T}_{\ell,j}(\epsilon)$
denotes the transmission function, which can be calculated by
evaluating the on-site retarded Green's function (GF) and lesser
Green's GF.[13] The indices ${\ell}$ and j denote the ${\ell}$th
QD and the $j$th QD, respectively. Based on the equation of motion
method, we can obtain analytical expressions of all GFs in the
Coulomb blockade regime. Details are provided in Ref. 13. The
transmission function in the weak interdot limit ( $t_c/U_{\ell} \ll
1$, where $t_c$ and $U_{\ell}$ denote the electron interdot hopping
strength and on-site Coulomb interaction, respectively) can be
recast into the following form
\begin{equation}
{\cal
T}_{\ell,j}(\epsilon)=-2\sum^{8}_{m=1}\frac{\Gamma_{\ell}(\epsilon)
\Gamma^{m}_{j}(\epsilon)}{\Gamma_{\ell}(\epsilon)+\Gamma^{m}_{j}(\epsilon)}
\mbox{Im}G^r_{\ell,m,\sigma}(\epsilon),
\end{equation}
where Im means taking the imaginary part of the function that
follows, and
\begin{equation}
G^r_{\ell,m,\sigma}(\epsilon)=p_m/(\mu_{\ell}-\Sigma_m).
\end{equation}
$\Gamma_{ \ell=L(1),R(2)}(\epsilon)$ denotes the tunnel rate from
the left electrode to dot A ($E_1$) and the right electrode to dot B
($E_2$), which is assumed to be energy- and bias-independent for
simplicity. $\mu_{\ell}=\epsilon-E_{\ell}+i\Gamma_{\ell}/2$. We can
assign the following physical meaning to Eq. (3). The sum in Eq. (3)
is over 8 possible configurations labeled by $m$. We consider an
electron (of spin $\sigma$) entering level $\ell$, which can be
either occupied (with probability $N_{\ell,\bar\sigma}$) or empty
(with probability $1-N_{\ell,\bar\sigma}$). For each case, the
electron can hop to level $j$, which can be empty (with probability
$a_j=1-N_{j,\sigma}-N_{j,\bar\sigma}+c_j$), singly occupied in a
spin $\bar\sigma$ state (with probability
$b_{j,\bar\sigma}=N_{j,\bar\sigma}-c_j$) or spin $\bar\sigma$ state
(with probability $b_{j,\sigma}=N_{j,\sigma}-c_j$), or a
double-occupied state (with probability $c_j$). Thus, the
probability factors associated with the 8 configurations appearing
in Eq. (4)  become $p_1=(1-N_{\ell,\bar\sigma})a_j$,
$p_2=(1-N_{\ell,\bar\sigma})b_{j,\bar\sigma}$,
$p_3=(1-N_{\ell,\bar\sigma})b_{j,\sigma}$,
$p_4=(1-N_{\ell,\bar\sigma})c_j$, $p_5=N_{\ell,\bar\sigma}a_j$,
$p_6=N_{\ell,\bar\sigma}b_{j,\bar\sigma}$,
$p_7=N_{\ell,\bar\sigma}b_{j,\sigma}$, and
$p_8=N_{\ell,\bar\sigma}c_j$. $\Sigma_m$ in the denominator of Eq.
(4) denotes the self-energy correction due to Coulomb interactions
and coupling with level $j$ (which couples with the other electrode)
in configuration $m$. We have $\Sigma_1=t^2_{c}/\mu_j$,
$\Sigma_2=U_{\ell,j}+t^2_{c}/(\mu_j-U_j)$,
$\Sigma_3=U_{\ell,j}+t^2_{c}/(\mu_j-U_{j,\ell})$,
$\Sigma_4=2U_{\ell,j}+t^2_{c}/(\mu_j-U_j-U_{j,\ell})$,
$\Sigma_5=U_{\ell}+t^2_{c}/(\mu_j-U_{j,\ell})$,
$\Sigma_6=U_{\ell}+U_{\ell,j}+t^2_{c}/(\mu_j-U_j-U_{j,\ell})$,
$\Sigma_7=U_{\ell}+U_{\ell,j}+t^2_{c}/(\mu_j-2U_{j,\ell})$, and
$\Sigma_8=U_{\ell}+2U_{\ell,j}+t^2_{c}/(\mu_j-U_j-2U_{j,\ell})$.
$E_{\ell}$, $U_{\ell}$, and $U_{\ell,j}$ denote, respectively, the
energy levels of dots, intradot Coulomb interactions, and interdot
Coulomb interactions. Here $\Gamma^m_{j}=-2$Im$ \Sigma_j$ denotes
the effective tunneling rate from level $l$ to the other electrode
through level $j$ in configuration $m$. For example,
$\Gamma^1_{j}=-2$Im$t^2_{c}/\mu_j=t^2_{c}\Gamma_j/[(\epsilon-E_j)^2+(\Gamma_j/2)^2]$.
It is noted that $\Gamma^m_{j}$ has a numerator $\Gamma_j$ for all
configurations. Furthermore,
$G^r_{\ell,\sigma}(\epsilon)=\sum^{8}_{m=1}
G^r_{\ell,m,\sigma}(\epsilon)$ is just the on-site single-particle
retarded GF for level ${\ell}$ as given in Eq. (A16) of Ref. 13, and
$G^r_{\ell,m,\sigma}(\epsilon)$ corresponds to its partial GF in
configuration $m$. The transmission function written this way has
the same form as Landauer's formula for a single QD with multiple
energy levels including intralevel and interlevel electron Coulomb
interactions.[14,15]

The probability factors of Eq. (3) are determined by the thermally
averaged one-particle occupation number and two-particle correlation
functions, which can be obtained by solving the on-site lesser
Green's functions,[13]

\begin{equation}
N_{\ell,\sigma}=-\int \frac{d\epsilon}{\pi}\sum^8_{m=1}
\frac{\Gamma_{\ell}f_{\ell}(\epsilon)+\Gamma^m_jf_j(\epsilon)}{\Gamma_{\ell}+\Gamma^m_{j}}
\mbox{Im}G^r_{\ell,m,\sigma}(\epsilon),
\end{equation}

and
\begin{equation}c_{\ell}=-\int
\frac{d\epsilon}{\pi}\sum^8_{m=5}
\frac{\Gamma_{\ell}f_{\ell}(\epsilon)+\Gamma^m_jf_j(\epsilon)}{\Gamma_{\ell}+\Gamma^m_{j}}
\mbox{Im}G^r_{\ell,m,\sigma}(\epsilon).
\end{equation}
Note that $\ell \neq j$ in Eqs. (3), (5), and (6). In the linear
response regime, Eqs. (1) and (2) can be rewritten as
\begin{eqnarray}
J&=&{\cal L}_{11} \frac{\Delta V}{T}+{\cal L}_{12} \frac{\Delta
T}{T^2}\\Q&=&{\cal L}_{21} \frac{\Delta V}{T}+{\cal L}_{22}
\frac{\Delta T}{T^2},
\end{eqnarray}
where $\Delta V=\mu_L-\mu_R$ and $\Delta T=T_L-T_R$ are the voltage
and temperature differences across the junction. Thermoelectric
response functions in Eqs. (7) and (8) are given by
\begin{equation}
{\cal L}_{11}=\frac{2e^2T}{h} \int d\epsilon {\cal T}(\epsilon)
(\frac{\partial f(\epsilon)}{\partial E_F})_T,
\end{equation}
\begin{equation}
{\cal L}_{12}=\frac{2eT^2}{h} \int d\epsilon {\cal T}(\epsilon)
(\frac{\partial f(\epsilon)}{\partial T})_{E_F},
\end{equation}

\begin{equation}
{\cal L}_{21}=\frac{2eT}{h} \int d\epsilon {\cal
T}(\epsilon)(\epsilon-E_F) (\frac{\partial f(\epsilon)}{\partial
E_F})_T,
\end{equation}
and
\begin{equation}
{\cal L}_{22}=\frac{2T^2}{h} \int d\epsilon {\cal T}(\epsilon)
(\epsilon-E_F)(\frac{\partial f(\epsilon)}{\partial T})_{E_F}.
\end{equation}
Here ${\cal T}(\epsilon)$ and
$f(\epsilon)=1/[e^{(\epsilon-E_F)/k_BT}+1]$ are evaluated in the
equilibrium condition. It can be shown that the Onsager relation
${\cal L}_{12}={\cal L}_{21}$ is preserved. These thermoelectric
response functions can also be found in ref. [7], where authors
investigated the thermoelectric properties of a single QD.

If the system is in an open circuit, the electrochemical potential
will form in response to a temperature gradient; this
electrochemical potential is known as the Seebeck voltage (Seebeck
effect). The Seebeck coefficient (amount of voltage generated per
unit temperature gradient) is defined as $S=\Delta V/\Delta T=-{\cal
L }_{12}/(T{\cal L}_{11})$. To judge whether the system is able to
generate power or refrigerate efficiently, we need to consider the
figure of merit,$^{1}$ which is given by
\begin{eqnarray}
ZT=\frac{S^2G_eT}{\kappa_e+\kappa_{ph}}\equiv \frac{(ZT)_0}{1+\kappa_{ph}/\kappa_e}.
\end{eqnarray}
Here $G_e={\cal L}_{11}/T$ is the electrical conductance and
$\kappa_e= (({\cal L}_{22}/T^2)-{\cal L}_{11}S^2)$ is the electron
thermal conductance. $(ZT)_0$ represents the $ZT$ value in the
absence of phonon thermal conductance, $\kappa_{ph}$. For
simplicity, we assume $\kappa_{ph}=\kappa_{ph,0} F_s$.[16-18]
$\kappa_{ph,0}=\frac{\pi^2k^2_B T}{3h}$ is the universal phonon
thermal conductance arising from acoustic phonon confinement in a
nanowire,[16-18] which was confirmed in the phonon wave guide.[19]
The expression of $\kappa_{ph}=\kappa_{ph,0} F_s$ with $F_s=0.1$ can
explain well the phonon thermal conductance of silicon nanowire with
surface states calculated by the first-principles method.[16] The
dimensionless scattering factor $F_s$ arises from phonon scattering
with surface impurities or surface defects of quantum dots.[1,16]
Here, we adopt $F_s$=0.02, which is smaller than $F_s$=0.1, because
QDs can enhance the phonon scattering rates and reduce phonon
thermal conduction as pointed out in Ref. 1.

\textbf{Results and discussion }

Here, we consider the case of identical QDs in the optimization of
$ZT$, although it is understood that the size fluctuation of QDs can
suppress $ZT$.[13] In Fig. 1(a) and 1(b), we plot $(ZT)_0$ and $ZT$
as a function of temperature for various electron hopping strengths.
We adopt the following physical parameters:
$E_{\ell}=E_F+30\Gamma_0$, $U_{\ell}=30\Gamma_0$,
$U_{\ell,j}=10\Gamma_0$, and $\Gamma_L=\Gamma_R=\Gamma=1\Gamma_0$.
All energy scales are in the units of the characteristic energy,
$\Gamma_0$. In Fig. 1(a), we see that $(ZT)_0$ increases with
decreasing $t_c$ and diverges as $t_c \rightarrow 0$. This behavior
can be proved rigorously as we shall illustrate below.  It implies
that SCQD can reach the Carnot efficiency in the limit of extremely
weak interdot coupling, if one can fully suppress $\kappa_{ph}$, for
example by inserting a nanoscale vacuum layer to block the phonon
heat current. Although it would be a challenging task to implement a
vacuum layer between one of the electrodes and SCQD, it may be
possible to test this idea out via a scanning tunneling microscopic
(STM) experiment by using a set-up as shown in the inset of Fig.
1(a). In Fig. 1(b), we see that $ZT$ is enhanced with increasing
$t_c$ until $t_c$ reaches $3\Gamma_0$, and it becomes reduced for
higher $t_c$.

The diverging behavior of $(ZT)_0$ with respect to $t_c$ is further
illustrated by the results of Fig. 2(d). The maximum ZT is
suppressed in the presence of $\kappa_{ph}$, which is much larger
than $\kappa_e$ for small $t_c$. The behaviors of ZT shown in Fig.
1(b) are mostly determined by the power factor $(S^2G_e$). Once
$t_c$ is larger than $3\Gamma_0$, the reduction of $S^2$ is faster
than the increase of $G_e$. This explains why the maximum $ZT$ at
$t_c=4\Gamma_0$ is smaller than that at $t_c=3\Gamma_0$. The
location of $ZT_{max}$ is nearly independent of $t_c$, and it occurs
near $k_BT=8.8\Gamma_0$. For comparison, we also show the results
(curves with triangle marks) for the case without electron Coulomb
interactions in Fig. 1(b). It is seen that the maximum $ZT$ is
enhanced when we turn off the electron Coulomb interactions. Such a
behavior is similar to that of a single QD with multiple energy
levels.[7,8] The effect of electron Coulomb interactions is
significant only for temperature between $6 \Gamma_0$ and
$50\Gamma_0$. Namely, the electron Coulomb interactions are
negligible when $U/(k_BT) \gg 1$ or $U/(k_BT) \ll 1$.

To further understand the behavior of $ZT$ with respect to $t_c$, we
plot the electrical conductance $(G_e)$, Seebeck coefficient ($S$),
electrical conductance $\kappa_e$, and $(ZT)_0$ as functions of
$t_c$ in Fig. 2 for various detuning energies, $\Delta\equiv
E_{\ell}-E_F$. When $E_{\ell}$ is close to the Fermi energy, $G_e$
and $\kappa_e$ are enhanced, whereas $S$ and $(ZT)_0$ are
suppressed. The behavior of  $(ZT)_0$ at $\Delta=30\Gamma_0$ in the
absence of Coulomb interactions is also shown by the curve with
triangles, which has similar trend as the solid line. Thus, it is
instructive to analyze $(ZT)_0$ in the absence of Coulomb
interactions. Keeping the leading order of $t^2_c$, we have ${\cal
L}_{11}=\frac{2e^2}{hk_B}
\frac{t^2_c}{\Gamma_0/2}\frac{1}{cosh^2(\Delta/2k_BT)}$, ${\cal
L}_{12}={\cal L}_{21}=\frac{2e}{hk_B}
\frac{t^2_c}{\Gamma_0/2}\frac{\Delta}{cosh^2(\Delta/2k_BT)}$,and
${\cal L}_{22}=\frac{2}{hk_B}
\frac{t^2_c}{\Gamma_0/2}\frac{\Delta^2}{cosh^2(\Delta/2k_BT)}$.
Therefore, $G_e \propto t^2_c$, $S=-k_B\Delta/eT$ is independent on
$t_c$, and $\kappa_e=({\cal L}_{22}-{\cal L}^2_{12}/{\cal
L}_{11})/T^2$ vanishes up to $t^2_c$. Thus, the leading order of
$\kappa_e$ is $t^4_c$. This indicates that $(ZT)_0 \propto 1/t^2_c$
in the limit of weak interdot hopping.

Figure 3 shows $ZT$ as a function of $\Delta=E_{\ell}-E_F$ for
various electron hoping strengths at $k_BT=10\Gamma_0$. Other
physical parameters are kept the same as those for Fig. 1. When
$t_c=0.1\Gamma_0$, the maximum $ZT$ ($ZT_{max}$) occurs at near
$\Delta=27\Gamma_0$. The peak position only  shifts  slightly to
higher $\Delta$ with increasing $t_c$. We have $ZT_{max}=2.79$ and
$3.18$ for  $t_c=1\Gamma_0$ and $3\Gamma_0$, respectively. However,
at $t_c=4\Gamma_0$ we have $ZT_{max}=3.07$, which is smaller than
$ZT_{max}$ for $t_c=3\Gamma_0$ . Thus, it also illustrates that $ZT$
is not a monotonically increasing function of $t_c$. We further
calculated $ZT$ as a function of $t_c$ for $\Delta=10,20,
30\Gamma_0$ and $k_BT=10\Gamma_0$ in the presence of $\kappa_{ph}$
and found that again ZT is not monotonically increasing function of
$t_c$ (not shown here). We conclude that  as long as $\kappa_{ph}$
dominates over $\kappa_e$, the $t_c$ dependence of ZT is mainly
determined by the power factor $S^2G_e$, where the behaviors of
$G_e$ and $S$ are similar to the results shown in Fig. 2(a) and
2(b).  When $t_c/\Gamma_0 \le 1$, $G_e$ increases much faster that
the reduction of $S^2$ for increasing  $t_c$, and the power factor
slowly reaches the maximum when $t_c$ approaches $3\Gamma_0$. When
$t_c > 3\Gamma_0$, the power factor decreases due to the fast
reduction of $S^2$ which prevails over the increase of $G_e$. The
curve with triangle marks is for $t_c=3\Gamma_0$ in the absence of
Coulomb interaction. We see that $ZT_{max}$ is larger when
$U_{\ell}=U_{\ell,j}=0$. Based on the results of Fig. 3, we conclude
that it is important to control the detuning energy $\Delta$ for the
optimization of $ZT$.

In Figures 1,2 and 3 we have considered the case with $E_F$ below QD
energy levels. It would be interesting to investigate the case with
$E_F$ above the energy levels of QDs. Fig. 4 shows $G_e$, S,
$\kappa_e$, and $ZT$ of an SCQD with $t_c=3\Gamma_0$ as functions of
applied gate voltage for various temperatures. Once $t_c >
(\Gamma_L+\Gamma_R)=2\Gamma_0$, the eight peaks for $G_e$ can be
resolved at $k_BT=1\Gamma_0$. These eight peaks correspond to the
following resonant channels: $E_{\ell}-t_c$, $E_{\ell}+t_c$,
$E_{\ell}+U_{\ell,j}-t_c$, $E_{\ell}+U_{\ell,j}+t_c$,
$E_{\ell}+U_{\ell,j}+U_{\ell}-t_c$,
$E_{\ell}+U_{\ell,j}+U_{\ell}+t_c$,
$E_{\ell}+2U_{\ell,j}+U_{\ell}-t_c$, and
$E_{\ell}+2U_{\ell,j}+U_{\ell}+t_c$, which are tuned by the gate
voltage to be aligned with $E_F$. These eight channels result from
the four configurations of $p_1$, $p_3$, $p_6$, and $p_8$ in Eq.
(4). Such a result implies that SCQD with identical QDs acts as a QD
with effective two levels of $E_{\ell}-t_c$ and $E_{\ell}+t_c$ and
satisfying Hund's rule. These eight peaks are smeared out with
increasing temperature. The sign changes of $S$ with respect to the
gate voltage result from the bipolar effect, i.e., the competition
between electrons and holes, where holes are defined as the
unoccupied states below $E_F$.[13] The electronic thermal
conductance ($\kappa_e$ ) also exhibits eight peaks, and we noticed
that the local maxima of the $\kappa_e$ curve nearly coincide with
the local minima of the $G_e$ curve. We see that $ZT$ values are
still larger than 3 even when $E_{\ell}$ is deeply below $E_F$ (say,
at $eV_g=70\Gamma_0$). This is attributed to the electron Coulomb
interaction. To illustrate that, we also show the results with
$U_{\ell}=U_{\ell,j}=0$ at $k_BT=3\Gamma_0$ (see the curve with
triangle marks). The oscillation of $ZT$ in the case of
$U_{\ell}=U_{\ell,j}=0$ is attributed to the sign change of $S$ at
$V_g=10\Gamma_0$. Note that $S$ goes to zero at $V_g=10\Gamma_0$,
which results from the electron-hole symmetry (with $E_{\ell}+t_c$
and $E_{\ell}-t_c$ straddling $E_F$ symmetrically). We see that $ZT$
vanishes for $eV_g \ge 40 \Gamma_0$ in the absence of electron
Coulomb interactions. Unlike the case of $E_F<E_{\ell}$, where the
finite $U$ causes reduction of $ZT$, here the electron Coulomb
interaction leads to enhancement of $ZT$ when $E_F>E_{\ell}$.

\textbf{Conclusions}

In summary, the thermoelectric properties including $G_e$, $S$,
$\kappa_e$ and $ZT$ of the SCQD junction system are investigated
theoretically.  We demonstrate that the Carnot efficiency can be
reached when $t_c$ approaches zero in the absence of phonon thermal
conductance. When the phonon contribution dominates the thermal
conductance of SCQD junction, the optimization of $ZT$ can be
obtained by the thermal power defined as $S^2 G_e$. We also found
that the presence of electron Coulomb interactions can lead to
either reduction or enhancement of $ZT$ depending on whether the
Fermi level is below or above the QD level.

{\bf Acknowledgment}\\
This work was supported in part by  National Science Council, Taiwan under Contract Nos. NSC 99-2112-M-008-018-MY2
and NSC 98-2112-M-001-022-MY3.
\mbox{}\\
$^{\dagger}$ E-mail address: mtkuo@ee.ncu.edu.tw\\
$^*$ E-mail address: yiachang@gate.sinica.edu.tw


\mbox{}\\
{\bf Figure Captions}

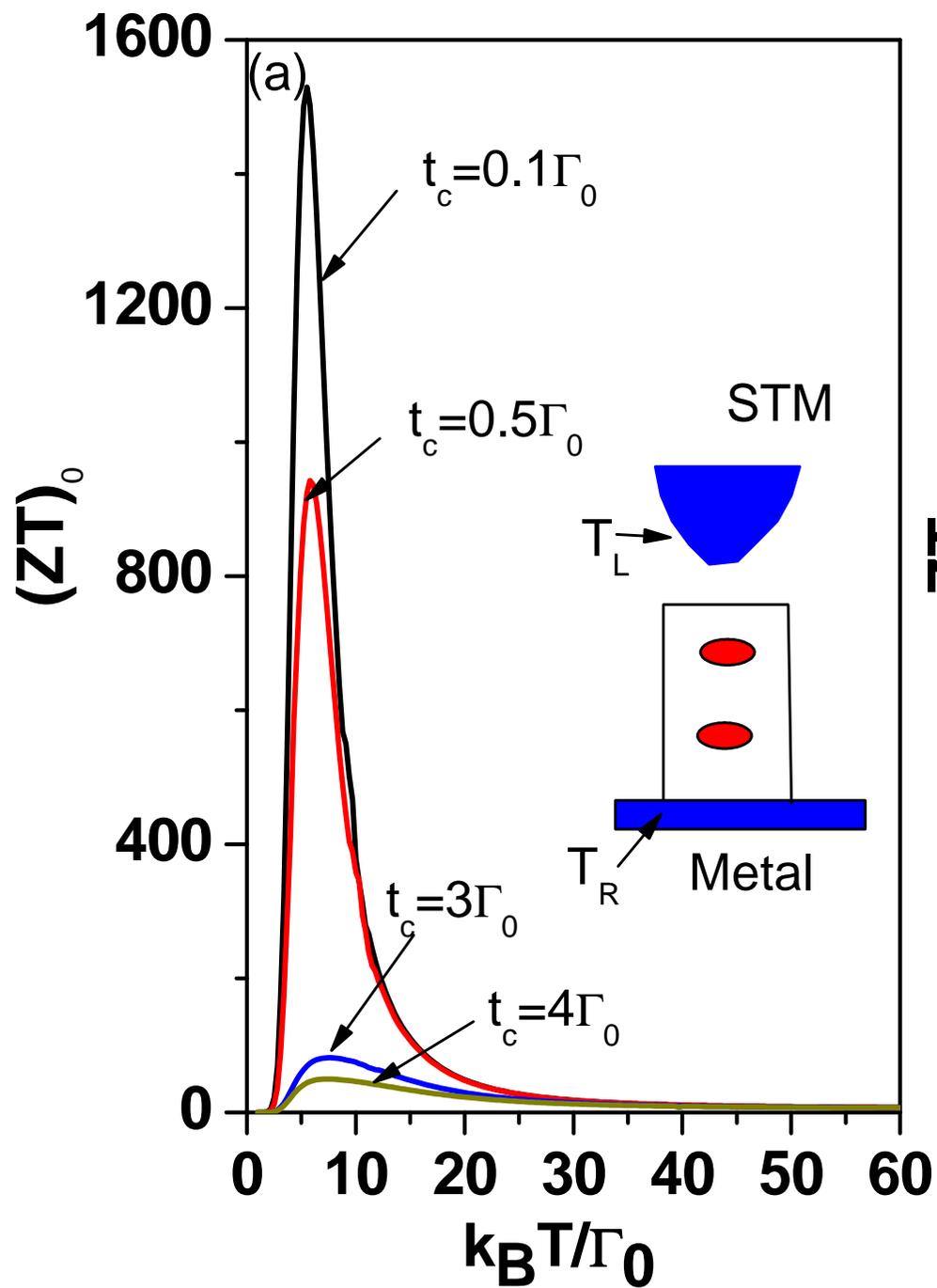
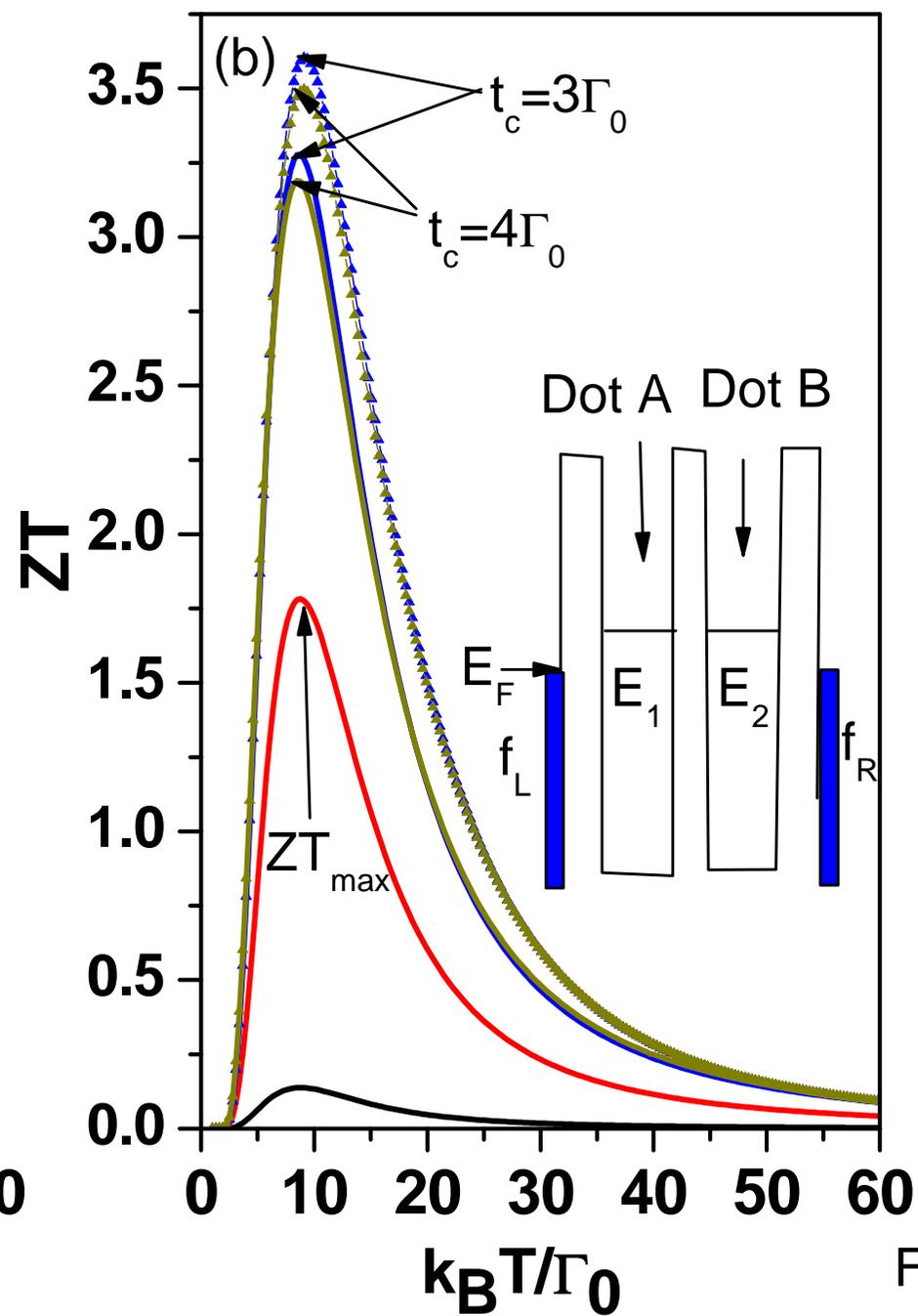
Fig. 1. (a) $(ZT)_0$ and (b) $ZT$ as functions of temperature for
various interdot hopping strengths ($t_c=0.1, 0.5, 1, 3$, and $4 \Gamma_0$).  $E_{\ell}=E_F+30\Gamma_0$,
$U_{\ell}=30\Gamma_0$, $U_{\ell,j}=10\Gamma_0$, and
$\Gamma_L=\Gamma=\Gamma_0$.


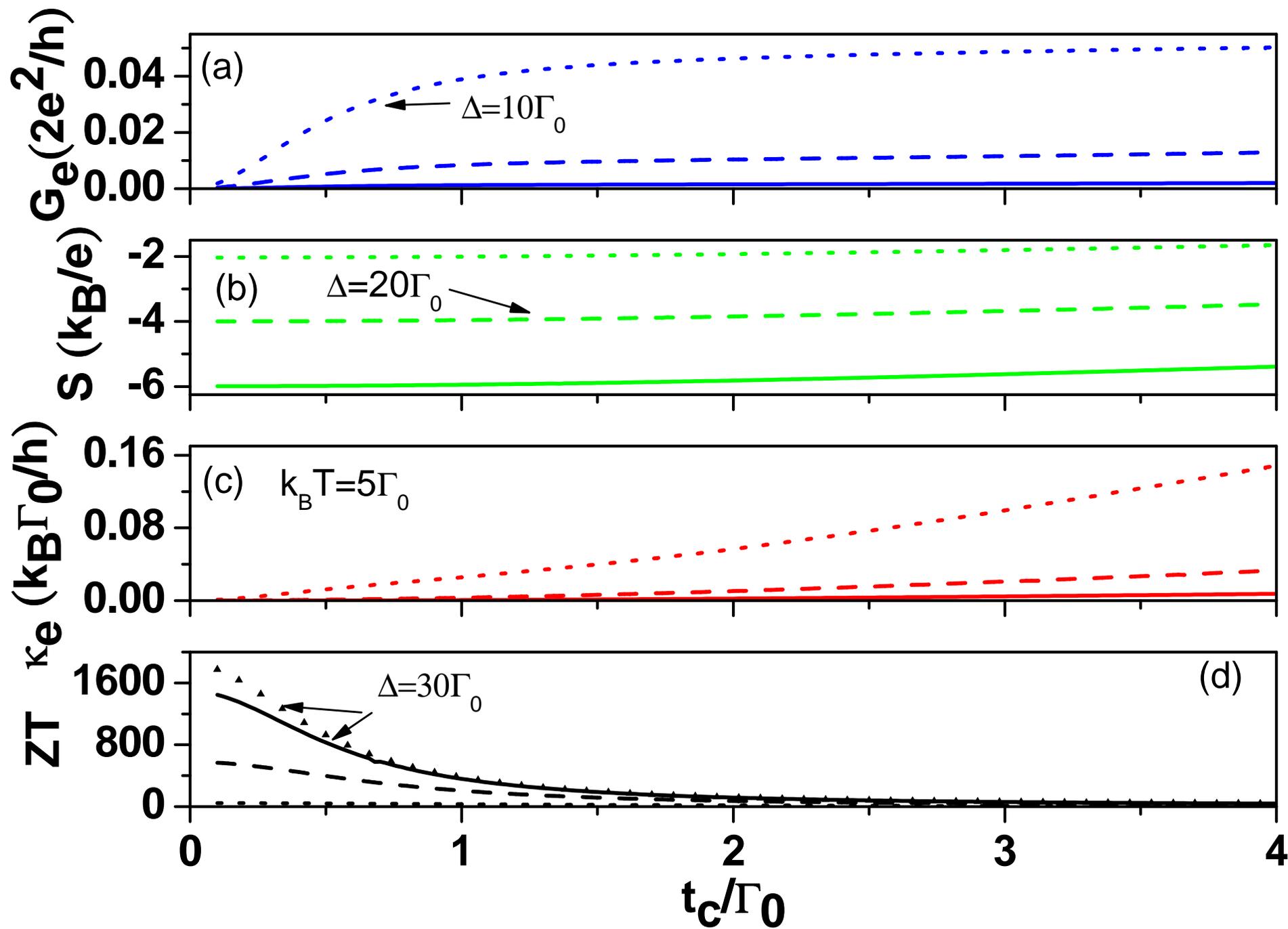
Fig. 2. Electrical conductance $(G_e)$, Seebeck coefficient (S),
electrical thermal conductance ($\kappa_e$), and (ZT)$_0$ as
functions of $t_c$ at $k_BT=5\Gamma_0$  for $\Delta=10\Gamma_0$ (dotted curves), $20\Gamma_0$ (dashed curves),
and $30\Gamma_0$ (solid curves). Other parameters are the same as those of
Fig. 1.

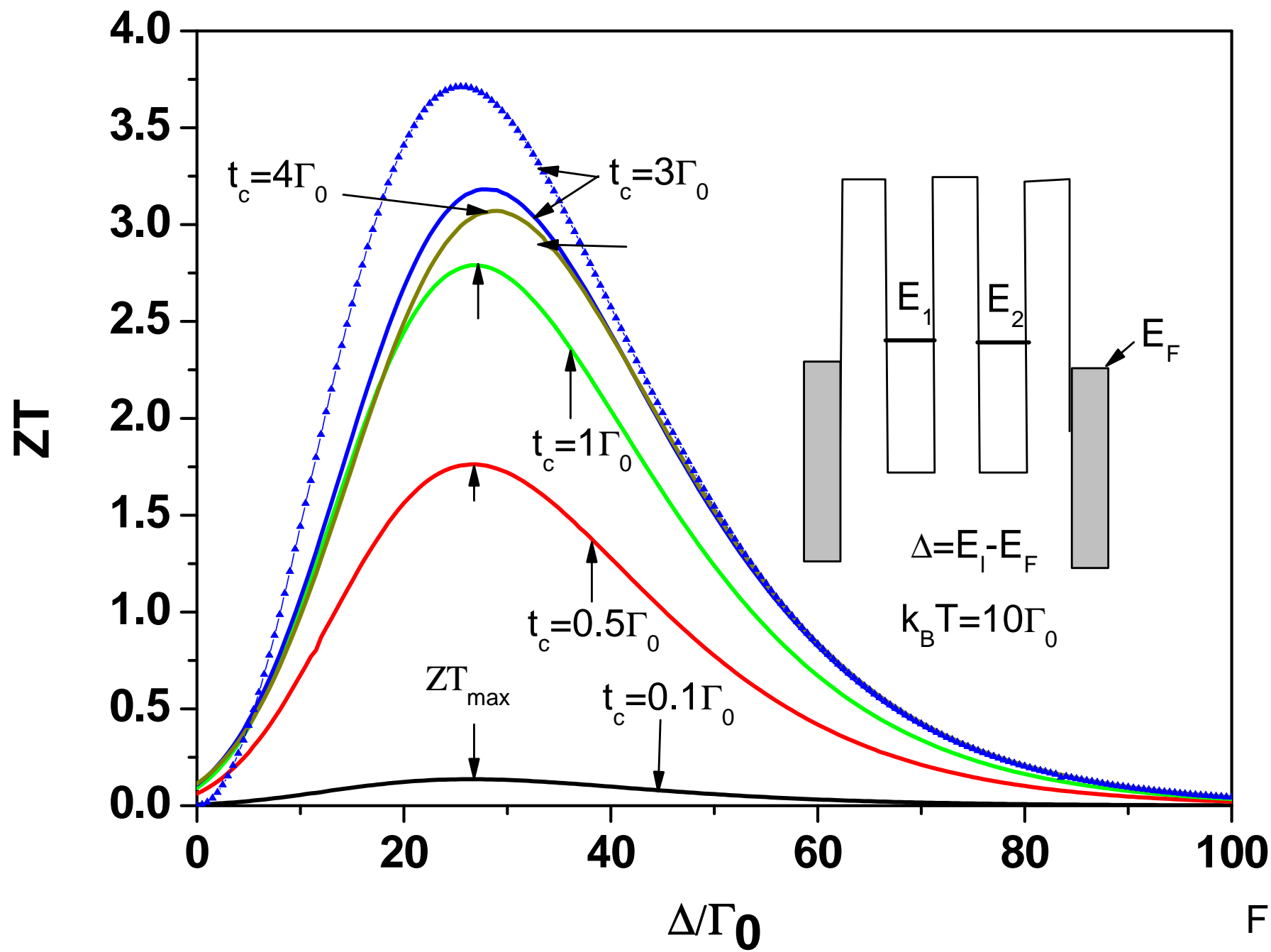
Fig. 3. $ZT$ as a function of $\Delta$ for different electron hopping
strength at $k_BT=10\Gamma_0$. Other parameters are the
same as those of Fig. 1.



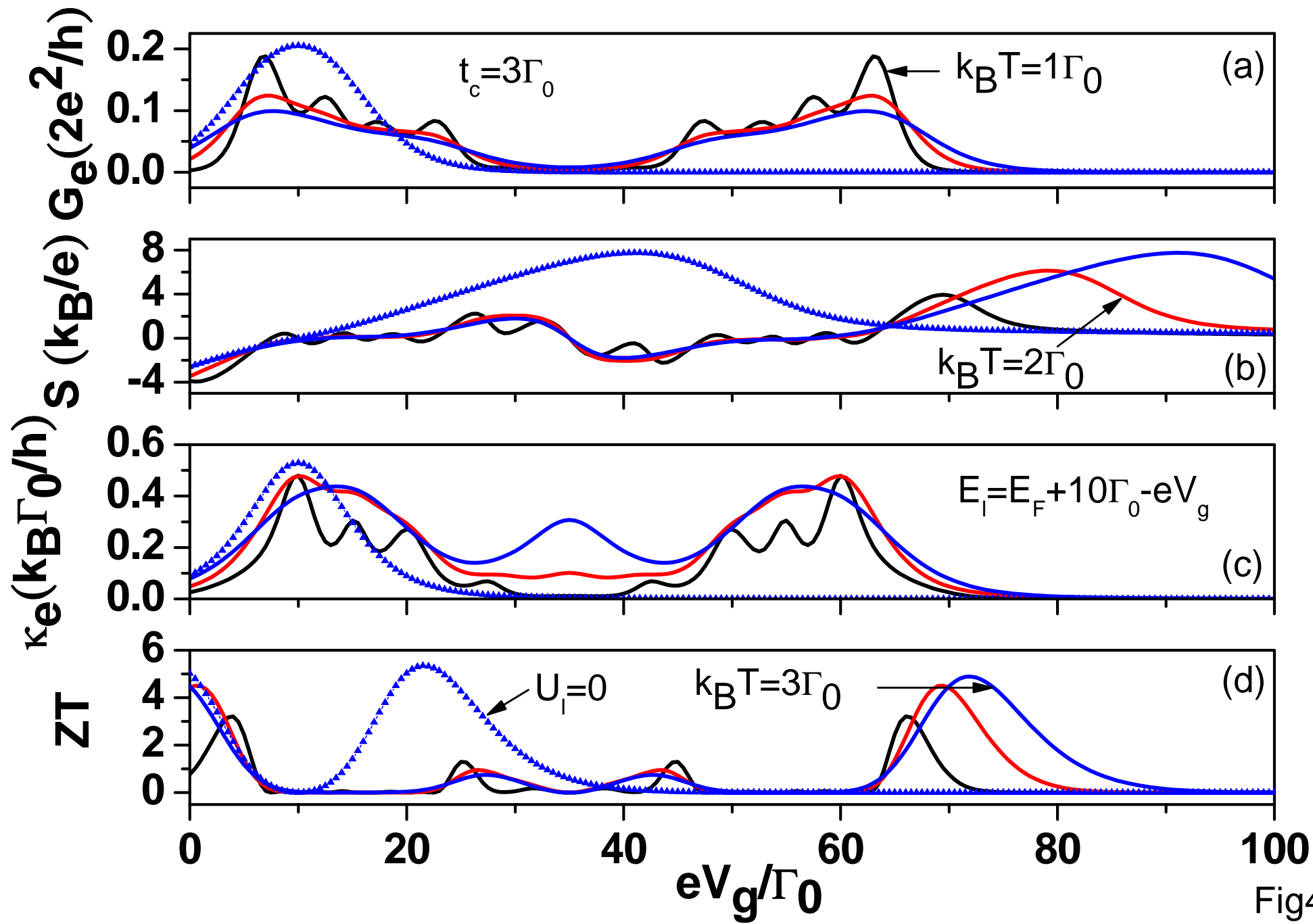
Fig. 4. $G_e$, S, $\kappa_e$, and $ZT$ as a function of applied gate
voltage for $k_BT=1 \Gamma_0$ (solid), $2 \Gamma_0$ (dashed), and  $3 \Gamma_0$ (dotted). $E_{\ell}=E_F+10\Gamma_0$ and
$t_c=3\Gamma_0$. Other parameters are the same as those of
Fig. 1. The curves with triangle marks are for the case without electron Coulomb interactions for
$k_BT=3\Gamma_0$ .


\begin{thebibliography}{50}

\bibitem[1]{Min} A. J. Minnich, M. S. Dresselhaus, Z. F. Ren,
G. Chen: \textbf{Bulk nanostructured thermoelectric materials:
current research and future prospects}, Energy Environ Sci 2009,
\textbf{2}:466-479.

\bibitem[2]{Mah} G. Mahan, B. Sales, J. Sharp: \textbf{Thermoelectric materials: New approaches to an old problem},
Physics Today 1997, \textbf{50}: (3) 42-47.

\bibitem[3]{Ven} R. Venkatasubramanian, E. Siivola, T. Colpitts, B. O'Quinn:
\textbf{Thin-film thermoelectric devices with high room-temperature
figures of merit}, Nature 2001,\textbf{413}: 597-602.

\bibitem[4]{Bou} A. I. Boukai, Y. Bunimovich, J. Tahir-Kheli, J. K.
Yu, W. A. Goddard III, J. R. Heath:\textbf{Silicon nanowires as
efficient thermoelectric materials }, Nature 2008, \textbf{451}:
168-171.

\bibitem[5]{Har} T. C. Harman, P. J. Taylor, M. P. Walsh, B. E.
LaForge:\textbf{Quantum dot superlattice thermoelectric materials
and devices}, Science 2002, \textbf{297}:  2229-2232.


\bibitem[6]{Hsu} K. F. Hsu, S. Loo, F. Guo,W. Chen, J. S. Dyck, C. Uher, T. Hogan,
E. K. Polychroniadis, M. G. Kanatzidis:\textbf{ Cubic AgPbmSbTe2+m:
Bulk thermoelectric materials with high figure of merit}, Science
2004,\textbf{303}: 818-821.

\bibitem[7]{Mur} P. Murphy, S. Mukerjee, J. Moore: \textbf{Optimal thermoelectric figure of merit of a molecular junction
}, Phys. Rev. B 2008,\textbf{78}: 161406-161410.


\bibitem[8]{Kuo} D. M. T. Kuo, Y. C. Chang:\textbf{ Thermoelectric and thermal rectification properties of quantum dot junctions},
Phys. Rev. B 2010,\textbf{ 81}: 205321-205331.


\bibitem[9]{Dub} Y. Dubi, M. Di Ventra:\textbf{ Heat flow and thermoelectricity in atomic and molecular junctions},
Rev. Modern Phys. 2011,\textbf{83}: 131-155.


\bibitem[10]{Ono} K. Ono, D. G. Austing, Y. Tokura, S. Tarucha:
\textbf{Current rectification by Pauli exclusion in a weakly coupled
double quantum dot system}, Science 2002, \textbf{297}: 1313-1317.



\bibitem[11]{Fra} J. Fransson, M. Rasander: \textbf{Pauli spin blockade in weakly coupled double quantum dots},
Phys. Rev. B 2006,\textbf{73}: 205333-205342.



\bibitem[12]{Sun}Q. F. Sun, Y. Xing, S. Q. Shen:
\textbf{Double quantum dot as detector of spin bias},Phys. Rev. B
2008, \textbf{77}: 195313.


\bibitem[13]{Kuo3} D. M. T. Kuo, S. Y. Shiau, Y. C. Chang:
\textbf{Theory of spin blockade, charge ratchet effect, and
thermoelectrical behavior in serially coupled quantum dot system},
Phys. Rev. B 2011, \textbf{84}: 245303-245314.

\bibitem[14]{Kuo4} D. M. T. Kuo, Y. C. Chang:
\textbf{Tunneling current spectroscopy of a nanostructure junction
involving multiple energy levels}, Phys. Rev. Lett. 2007,
\textbf{99}: 086803-086807.

\bibitem[15]{cha}  Y. C. Chang, D. M. T. Kuo:
\textbf{Theory of charge transport in a quantum dot tunnel junction
with multiple energy levels}, Phys. Rev. B 2008, \textbf{77}:
245412-245428.


\bibitem[16]{Mar} T. Markussen, A. P. Jauho, M. Brandbyge:
\textbf{Surface-Decorated Silicon Nanowires: A Route to High-ZT Thermoelectrics},
 Phys. Rev. Lett. 2009,\textbf{ 103}: 055502-055506.


\bibitem[17]{San1} D. H. Santamore, M. C. Cross:
\textbf{Effect of phonon scattering by surface roughness on the
universal thermal conductance }, Phys. Rev. Lett. 2001, \textbf{87}:
115502-115506.



\bibitem[18]{Reg} L. G. C. Rego, G. Kirczenow:
\textbf{Quantized thermal conductance of dielectric quantum wires},
Phys. Rev. Lett. 1998,\textbf{81}: 232-236.


\bibitem[19]{Sch} K. Schwab, E. A. Henriksen, J. M. Worlock,M.
L. Roukes:\textbf{Measurement of the quantum of thermal conductance}
,Nature 2000, \textbf{404}: 974-977.









\end{thebibliography}
\end{document}